\documentclass[dvipdfmx,aps,prd,showpacs,preprintnumbers,superscriptaddress,nofootinbib,twocolumn]{revtex4}%
\usepackage[dvipdfmx]{graphicx}
\usepackage{bm,latexsym,amsmath,amssymb,amsfonts,mathrsfs,subfigure}
\usepackage{color}
\input{colordvi.tex}
\usepackage[dvipdfmx]{hyperref}
\usepackage{url}
\hypersetup{
    colorlinks=true,
    citecolor=cyan,
}
\newcommand*{\mpl}{M_{\rm Pl}}
\newcommand*{\dd}{\text{\large$\cdot$}}
\begin{document}

\title{Matter bispectrum beyond Horndeski}

%%%%%%%%%%%%%%%%%%%%%%%%%%%%%%%%%%%%%%%%%%%%%%%%%

\author{Shin'ichi Hirano}
\email[Email: ]{s.hirano"at"rikkyo.ac.jp}
\affiliation{Department of Physics, Rikkyo University, Toshima, Tokyo 171-8501, Japan}

\author{Tsutomu Kobayashi}
\email[Email: ]{tsutomu"at"rikkyo.ac.jp}
\affiliation{Department of Physics, Rikkyo University, Toshima, Tokyo 171-8501, Japan}

\author{Hiroyuki Tashiro}
\email[Email: ]{hiroyuki.tashiro"at"nagoya-u.jp}
\affiliation{Department of Physics, Graduate School of Science, Nagoya University, Aichi 464-8602, Japan}

\author{Shuichiro Yokoyama}
\email[Email: ]{shuichiro"at"rikkyo.ac.jp}
\affiliation{Department of Physics, Rikkyo University, Toshima, Tokyo 171-8501, Japan}
\affiliation{Kavli IPMU (WPI), UTIAS, The University of  Tokyo, Kashiwa, Chiba 277-8583, Japan}

%%%%%%%%%%%%%%%%%%%%%%%%%%%%%%%%%%%%%%%%%%%%%%%%%

\begin{abstract}
The Horndeski scalar-tensor theory and its recent extensions allow nonlinear derivative interactions of the scalar degree of freedom.
We study the matter bispectrum of large scale structure as a probe of these modified gravity theories,
focusing in particular on the effect of the terms that newly appear in the so-called ``beyond Horndeski'' theories.
We derive the second-order solution for the matter density perturbations and find that the interactions beyond Horndeski lead to a new time-dependent coefficient in the second-order kernel which differs in general from the standard value of general relativity and the Horndeski theory.
This can deform the matter bispectrum at the folded triangle configurations ($k_1+k_2=k_3$),
while it is never possible within the Horndeski theory.
\end{abstract}

\pacs{04.50.Kd,~98.65.-r,~98.80.Cq}
\preprint{RUP-18-3}
\maketitle
%%%%%%%%%%%%%%%%%%%%%%%%%%%%%%%%%%%%%%%%%%%%%%%%%%%%%%%%%%%%%%%%%%%%%%

\section{Introduction}

The discovery of the accelerated expansion of the universe~\cite{Riess:1998cb, Perlmutter:1998np} has stimulated exploring consistent modification of gravity as an alternative to the cosmological constant or dark energy (see, {\em e.g.}, Refs.~\cite{Clifton:2011jh,Joyce:2014kja,Koyama:2015vza,Bull:2015stt} for a review).
On the largest scale, the evolution of the homogeneous background of the conventional $\Lambda$CDM model is supposed to be mimicked by modification of gravity, and therefore the accelerated expansion driven by modified gravity is indistinguishable by definition from that of the standard scenario.
On small scales inside the solar system, the effect of modification of gravity must be highly suppressed, as we have stringent constraints on the deviation from general relativity~\cite{Will:2014kxa}.
It is not difficult for scalar-tensor theories, by which a large class of modified gravity can be described at least effectively, to incorporate the mechanism such as the Vainshtein mechanism~\cite{Vainshtein:1972sx} to hide the force mediated by the additional scalar degree of freedom in the solar system.
Going to cosmological scales in between, we have an intriguing arena for tests of gravity offered by large scale structure of the universe through its linear and nonlinear evolution.

The Horndeski theory~\cite{Horndeski:1974wa, Deffayet:2011gz, Kobayashi:2011nu} is the most general scalar-tensor theory having second-order equations of motion, and this is a useful framework with which to study modified gravity models in a comprehensive way.
Since it shares the same structure of nonlinear derivative interaction as the Galileon theory~\cite{Nicolis:2008in}, the Vainshtein screening mechanism can naturally be implemented~\cite{Kimura:2011dc, Narikawa:2013pjr, Koyama:2013paa}.
It is expected that this derivative nonlinearity is imprinted in the one-loop dark-matter power spectrum and the bispectrum.
This point has been investigated within the Horndeski theory in Refs.~\cite{Takushima:2013foa, Bellini:2015wfa, Cusin:2017wjg}.
Recently, it was noticed that the Horndeski theory can further be generalized while maintaining the number of propagating degrees of freedom (one scalar and two tensor modes)~\cite{Zumalacarregui:2013pma,Gleyzes:2014dya, Gleyzes:2014qga, Langlois:2015cwa, Crisostomi:2016czh, Achour:2016rkg, BenAchour:2016fzp}.
Such theories are called {\em degenerate} higher-order scalar-tensor theories.
Higher derivative terms in the equations of motion of degenerate theories disappear in the end after combining different components, and hence such theories are Ostrogradsky-stable~\cite{Ostro1, Ostro2}.
New operators arise in degenerate scalar-tensor theories beyond Horndeski,
and one of the interesting effects due to them is the partial breaking of Vainshtein screening inside matter~\cite{Kobayashi:2014ida}. 
These new derivative interactions will also participate in the one-loop matter power spectrum
and the bispectrum, which could be a probe of modified gravity theories beyond Horndeski.
See Refs.\cite{Koyama:2015oma,Saito:2015fza,Sakstein:2015zoa,Sakstein:2015aac,
Jain:2015edg,Babichev:2016jom,Sakstein:2016lyj,Sakstein:2016oel,Sakstein:2016ggl,Dima:2017pwp} for other probes of degenerate higher-order scalar-tensor theories.

The purpose of this paper is to investigate the impact of the new operators of the Gleyzes-Langlois-Piazza-Vernizzi (GLPV) theory~\cite{Gleyzes:2014dya, Gleyzes:2014qga} on the matter bispectrum.
As the GLPV theory (without the so-called $F_5$ term) is the simplest extension of the Horndeski theory in the context of degenerate theories, this work is a first step to study how new nonlinear interactions beyond Horndeski affect non-Gaussianity of large scale structure.

This paper is organized as follows.
In the next section, we derive our basic equations for the matter density perturbations $\delta$ in the GLPV theory.
We then give a second-order solution for $\delta$ in Sec.~III.
In Sec.~IV the matter bispectrum in the GLPV theory is evaluated
and its particular feature is emphasized.  
In Sec.~V, we give a short comment on the implication of the recent gravitational wave event
for the theory considered in the present paper.
We draw our conclusions in Sec.~VI.

\newpage
\section{Basic Equations}

\subsection{The GLPV theory}

The action of the GLPV theory is given by~\cite{Gleyzes:2014dya, Gleyzes:2014qga}
\begin{align}
     S = \int d^{4}x \sqrt{-g} \left( {\cal L}  + {\cal L}_{\rm m}
 \right)\label{theory},
\end{align}
where\footnote{Concerning the factors in front of $F_4$ and $F_5$,
 we follow the convention of Ref.~\cite{Kobayashi:2014ida}
 which is different from the one used in Ref.~\cite{Gleyzes:2014dya,Gleyzes:2014qga}.}
\begin{align}
    {\cal L} 
        & = G_{2}(\phi,X)\ -G_{3}(\phi,X)\Box\phi\ 
            \notag\\
        & \quad + G_{4}(\phi,X)R + G_{4X} \left[(\Box\phi)^{2}-\phi_{\mu\nu}^{2} \right] 
            \notag\\
        & \quad +G_{5}(\phi,X)G_{\mu\nu}\phi^{\mu\nu} -\frac{1}{6}G_{5X} 
            \notag\\
        & \quad\quad\times [(\Box\phi)^{3} -3(\Box\phi)\phi^{2}_{\mu\nu} +2\phi^{3}_{\mu\nu}] 
            \notag\\
        & \quad -\frac{1}{2}F_{4}(\phi,X) \epsilon^{\mu\nu\rho\sigma} \epsilon_{\mu'\nu'\rho'\sigma}
            \phi^{\mu'}\phi_{\mu}\phi^{\nu'}_{\ \nu}\phi^{\rho'}_{\ \rho} 
                \notag\\
        & \quad -\frac{1}{3}F_{5}(\phi, X) \epsilon^{\mu\nu\rho\sigma} \epsilon_{\mu'\nu'\rho'\sigma'}
            \phi^{\mu'}\phi_{\mu}\phi^{\nu'}_{\ \nu}\phi^{\rho'}_{\ \rho}\phi^{\sigma'}_{\ \sigma}, 
                \label{eq: GLPV_Lag}
\end{align}
and ${\cal L}_{\rm m}$ is the Lagrangian of the matter components.
Here we use the notation $\phi_\mu:=\nabla_\mu\phi$,
$\phi_{\mu\nu}:=\nabla_\mu\nabla_\nu\phi$,
$G_X:=\partial G/\partial X$,
and $\epsilon^{\mu\nu\rho\sigma}$ is the totally antisymmetric Levi-Civita tensor.
The above Lagrangian has six arbitrary functions, 
$G_{i}\ (i=2,3,4,5)$ and $F_{j}\ (j=4,5)$,
of $\phi$ and $X:=(-1/2)\phi_\mu\phi^\mu$.
The GLPV theory is an extension of the Horndeski theory,
and Eq.~(\ref{eq: GLPV_Lag}) reduces to the Horndeski Lagrangian in the case of $F_4=F_5=0$.

Among wide classes of theories described by the GLPV action,
we focus on those with $G_5=F_5=0$ in the present paper.
This is a reasonable restriction because the $G_5$ term not only hinders the recovery of
the Newtonian behavior of the gravitational potentials on small scales in a cosmological background~\cite{Kimura:2011dc}, but also causes some instabilities insider the Vainshtein radius~\cite{Koyama:2013paa}.
Since the $F_5$ term has the structure similar to the $G_5$ term,
the same pathologies are expected, though this has not been confirmed explicitly so far.
In the absence of $G_5$ and $F_5$, the GLPV theory is degenerate without further conditions~\cite{Crisostomi:2016czh}, so that there are at most 3 propagating degrees of freedom in any background spacetime.
This nature is desirable in view of Ostrogradsky instabilities.

One of the interesting consequences of the $F_4$ term is the partial breaking of the Vainshtein screening mechanism inside matter sources~\cite{Kobayashi:2014ida, Koyama:2015oma, Saito:2015fza, Sakstein:2015zoa, Sakstein:2015aac, Jain:2015edg, Babichev:2016jom, Sakstein:2016lyj, Sakstein:2016oel, Sakstein:2016ggl, Dima:2017pwp},
where derivative nonlinearities are significant.
It turns out that the partial breaking of the Vainshtein mechanism generically occurs in degenerate higher-order scalar-tensor theories~\cite{Crisostomi:2017lbg, Langlois:2017dyl, Dima:2017pwp, Crisostomi:2017pjs}.
In the present paper, we study the impact of the nonlinearities of the $F_4$ term on the matter bispectrum.
Some studies in this direction have already been undertaken
in the context of the Horndeski theory in Refs.~\cite{Takushima:2013foa, Bellini:2015wfa, Cusin:2017wjg},
and this work is an extension of~\cite{Takushima:2013foa}.

\subsection{Effective action under the quasi-static approximation}

We consider cosmological perturbations in a homogeneous and isotropic cosmological background.
The field equations governing the background evolution are found in Ref.~\cite{Kobayashi:2014ida}.
As we are not interested in the evolution of the universe in a particular modified gravity model,
here we simply assume that the field equations admit a solution that is very close to the usual $\Lambda$CDM model.
This is in principle possible because we have the four free functions in the theory that can be tuned if necessary.

The perturbed metric in the Newtonian gauge is given by
\begin{align}
    ds^{2} = -(1+2\Phi)dt^{2} + a^{2}(t)(1-2\Psi)d{\bm x}^{2},
\end{align}
and the perturbed scalar field is written as
\begin{align}
    \phi (t,{\bm x})= \bar\phi(t) + \pi(t,{\bm x}),
\end{align}
where a barred variable denotes the background quantity.
It is convenient to introduce the dimensionless scalar field perturbation as
$Q(t,{\bm x}):=H\pi/\dot{\bar\phi}$.
We only consider nonrelativistic matter and write its energy density as
\begin{align}
    \rho_{\rm m} (t,{\bm x})= \bar\rho_{\rm m}(t)[1+\delta(t,{\bm x})],
\end{align}
where
$\delta$ is a density contrast.
In what follows we omit bars from the background quantities.

We expand the action~(\ref{theory}) in terms of the perturbations.
Since we are interested in the evolution of the density perturbations inside the (sound) horizon, we employ the quasi-static approximation,\footnote{The validity of the quasi-static approximation has been discussed in Refs.~\cite{Sawicki:2015zya, DAmico:2016ntq, DeFelice:2015isa}.
See also Refs.~\cite{Noller:2013wca,Chiu:2015voa,Winther:2015pta}.}
$\nabla_i\epsilon \gg \dot\epsilon\sim H\epsilon$, where
$\nabla_i$ is the spatial derivative, a dot stands for the time derivative, and
$\epsilon$ is any of $\Phi$, $\Psi$, or $\pi$.
This does not mean to drop all the time derivatives and the Hubble parameter,
because one may expect that
$\nabla^2\Phi/a^2\sim H^2 \delta\sim H\dot\delta\sim \ddot\delta$ 
and hence the time derivatives acting on $\delta$ cannot be ignored in general.
In the case of the GLPV theory, we will also have terms like
$\nabla^2\dot\Psi$ in the perturbation equations, which must be retained as well.

The crucial point in the perturbative expansion is that, in the Horndeski and GLPV theories,
the second derivatives of perturbations can be large on small scales even though the first and zeroth derivatives are small, 
so that the terms nonlinear in the second derivatives cannot be neglected.
This is the very reason why the Vainshtein screening mechanism (partially) works.
This is also the key nonlinearity for the matter bispectrum.

Noting that the matter Lagrangian can be written as
${\cal L}_{\rm m}=-\Phi \rho_{\rm m}\delta$,
we have the following effective action governing the
perturbation evolution in the quasi-static regime~\cite{Kobayashi:2014ida}:
\begin{align}
    S_{\mathrm{eff}} = \int dtd^{3}x \, a^{3}\, \left[ {\cal L}^{(2)} + {\cal L}^{\mathrm{(NL)}}\right],
\label{effctact}
\end{align}
where
\begin{align}
    {\cal L}^{\mathrm{(2)}}
    & = -M^{2}(1+\alpha_{T})\Psi\frac{\nabla^{2}\Psi}{a^{2}} 
        + 2M^{2}(1+\alpha_{H})\Psi\frac{\nabla^{2}\Phi}{a^{2}} 
            \notag\\
    & \quad  -M^{2}\biggl[\frac{\dot H}{H^{2}} +\frac{3\Omega_{\rm m}}{2} 
        +\biggl(1+\alpha_{M}+\frac{\dot H}{H^{2}}\biggr)(\alpha_{B}-\alpha_{H}) 
            \notag\\
    & \quad  +\frac{\dot\alpha_{B} -\dot\alpha_{H}}{H} 
        +(\alpha_{T} -\alpha_{M}) \biggr]Q\frac{\nabla^{2}Q}{a^{2}} 
            \notag\\
    & \quad -2M^{2}(\alpha_{B}-\alpha_{H})\Phi\frac{\nabla^{2}Q}{a^{2}} 
        \notag\\
    & \quad +2M^{2}\left[\alpha_{H}(1+\alpha_{M}) +\alpha_{M} -\alpha_{T}
        +\frac{\dot\alpha_{H}}{H}\right]\Psi\frac{\nabla^{2}Q}{a^{2}} 
            \notag\\
    & \quad  -\rho_{\mathrm{m}}\Phi\delta +2M^{2}\alpha_{H}\frac{\dot\Psi}{H}\frac{\nabla^{2}Q}{a^{2}},
        \label{L2}
            \\
    {\rm and~~\,}&
        \notag\\
    {\cal L}^{\mathrm{(NL)}} 
    & = \frac{M^{2}}{2H^{2}}\biggl[\alpha_{G} -3(\alpha_{H}-\alpha_{T}) +4\alpha_{B} 
        \notag\\
    & \quad -\alpha_{M}(2 +\alpha_{G} +\alpha_{H}) -\frac{\dot\alpha_{G} 
        +\dot\alpha_{H}}{H} \biggr]\frac{{\cal L}_{3}}{a^{4}} 
            \notag\\
    & \quad +\frac{M^{2}}{2H^{2}}(\alpha_{G} -\alpha_{H})\Phi\frac{{\cal Q}^{(2)}}{a^{4}} 
        +\frac{M^{2}}{2H^{2}}\alpha_{T}\Psi\frac{{\cal Q}^{(2)}}{a^{4}} 
            \notag\\
    & \quad -\frac{2M^{2}}{H^{2}}\alpha_{H}\frac{\nabla_{i}\Psi\nabla_{j}Q\nabla^{i}\nabla^{j}Q}{a^{4}}
        \notag\\
    & \quad + \frac{M^{2}}{2H^{4}}(\alpha_{G} -\alpha_{H} +\alpha_{T})\frac{{\cal L}_{4}}{a^{6}},
        \label{LN}
\end{align}
with
\begin{align}
    {\cal L}_{3} & = -\frac{1}{2}(\nabla Q)^{2}\nabla^{2}Q, 
        \\
    {\cal L}_{4} & = -\frac{1}{2}(\nabla Q)^{2}{\cal Q}^{(2)}, 
        \\
    {\cal Q}^{(2)} & = (\nabla^{2}Q)^{2} -(\nabla_{i}\nabla_{j}Q)^{2}.
\end{align}
The time-dependent parameters in the coefficients are
defined by
\begin{align}
    M^{2} & = 2(G_{4} -2XG_{4X} -2X^{2}F_{4}),
        \\
    \alpha_{M} & =H^{-1}\frac{d\ln M^2}{d t}, 
        \\
    HM^{2}\alpha_{B} 
        & = -\dot\phi (XG_{3X} -G_{4\phi} -2XG_{4\phi X}) -4HX 
            \notag\\
        & \quad \times(G_{4X} +2XG_{4XX} +4XF_{4} +2X^{2}F_{4X}), 
            \\
    M^{2}\alpha_{T} & = 4X (G_{4X} +XF_{4}), 
        \\
    M^{2}\alpha_{H} & = 4X^{2}F_{4},
\end{align}
and
\begin{align}
    \Omega_{\rm m} := \frac{\rho_{\rm m}}{3M^{2}H^{2}} 
        \label{fraction parameters},
\end{align}
which were introduced and used in Refs.~\cite{Bellini:2014fua, Gleyzes:2014qga, Gleyzes:2014rba, Ade:2015rim}.
(we follow the convention of Ref.~\cite{Gleyzes:2014qga}.)
We have defined another useful parameter as
\begin{align}
    M^{2}\alpha_{G} = 4X(G_{4X} +2XG_{4XX} +4XF_{4} +2X^{2}F_{4X}),
\end{align}
which first appears in the cubic order action.

The physical meanings of those parameters are as follows:
$M$ is the effective Planck mass, $\alpha_M$ is its evolution rate,
$\alpha_B$ is the braiding parameter that characterizes
the kinetic mixing of the scalar field and the metric, and $\alpha_T$ parameterizes the deviation of the speed of gravitational waves from that of light.
The $\alpha_H$ parameter signals novel effects compared to the Horndeski theory.
The last term in Eq.~(\ref{L2}) and the fourth line in ~Eq. (\ref{LN}), which generate third-order derivatives in the equations of motion, are proportional solely to this parameter and hence appear for the first time in the GLPV theory.
Note that $\Omega_{\rm m}$ cannot always be interpreted as the familiar density parameter,
because the Friedmann equation is modified and we do not necessarily have the equation of the form $3M^2H^2=\rho_{\rm m}$ $+$ the energy density of the scalar field.
This is related to the fact that the distinction between the geometry
(the ``left hand side'' of the gravitational field equations)
and the energy-momentum tensor is ambiguous in the presence of nonminimal coupling.

If all the $\alpha$ parameters vanish and $M=M_{\rm Pl}$ (the Planck mass),
the nonlinear part of the Lagrangian, ${\cal L}^{{\rm (NL)}}$, vanishes
and the quadratic Lagrangian ${\cal L}^{(2)}$ reduces to the standard expression in general relativity.
In view of this, we assume that
\begin{align}
    \alpha_M,\alpha_B,\alpha_T,\alpha_H,\alpha_G \ll 1,
\end{align}
in the early stage of the matter-dominant universe, so that standard cosmology is recovered.
In the late-time universe, however, the effect of modification of gravity emerges, which is assumed to be responsible for the accelerated expansion.
In this stage we assume ${\cal O}(1)$ modification from general relativity,
{\em i.e.},
\begin{align}
    \alpha_M,\alpha_B,\alpha_T,\alpha_H,\alpha_G ={\cal O}( 1).
\end{align}
This is equivalent to assuming that
\begin{align}
    & \dot\phi\sim \mpl H_{0},\ G_{2}\sim \mpl^{2}H_{0}^{2},~G_{3X}\sim \mpl^{-1}H_0^{-2}, 
        \notag\\
    & G_{4}\sim \mpl^{2},\ F_{4}\sim \mpl^{-2}H^{-4}_{0},~\cdots 
        \label{condition_{late}}
\end{align}
in the late-time universe, where the Hubble parameter is roughly given by its present value, $H_0$.

\subsection{Field equations in Fourier space}

Now we move to the field equations that can be derived by varying the effective action~(\ref{effctact}) with respect to
$\Psi, \Phi$, and $Q$.
They are given, in Fourier space,\footnotemark[3] by
\begin{widetext}
\begin{align}
     &-p^{2}\left[{\cal F}_{T}\Psi(t,{\bm p})-{\cal G}_{T}\Phi(t,{\bm p}) - A_{3}Q(t,{\bm p}) 
        + M^2\alpha_H \frac{\dot Q(t,{\bm p})}{H}\right]
            =\frac{B_{1}}{2a^{2}H^{2}}\Gamma[t,{\bm p};Q,Q]\notag \\
    &\hspace{12em}%~~~~~~~~~~~~~~~~~~~~~~~~~~~~~~~~~~~~~~~~~~~~~
    + \frac{M^2\alpha_H}{a^{2}H^{2}}\frac{1}{(2\pi)^{3}} \int d^{3}k_{1}d^{3}k_{2}\,
        \delta^{(3)}({\bm k}_{1}+{\bm k}_{2}-{\bm p}) k_{1}^{2}k_{2}^{2}\, 
            \beta({\bm k}_{1},{\bm k}_{2}) Q(t,{\bm k}_{1})Q(t,{\bm k}_{2}),
                \label{Psi F} \\
%%%%%%%%%%%%%%%%%%%%
    &\hspace{7em}
    -p^{2}\left[ {\cal G}_{T}\Psi(t,{\bm p})+A_{2}Q(t,{\bm p})\right]  -\frac{a^{2}}{2}\rho_{\rm m}\delta(t,{\bm p})
        = -\frac{B_{2}}{2a^{2}H^{2}}\Gamma[t,{\bm p};Q,Q],
            \label{Phi F}\\
%%%%%%%%%%%%%%%%%%%%
    &-p^{2} \left[ A_{0}Q(t,{\bm p}) -A_{1}\Psi(t,{\bm p}) -A_{2}\Phi(t,{\bm p}) 
        -M^2\alpha_H\frac{\dot\Psi(t,{\bm p})}{H}\right]
            = -\frac{B_{0}}{a^{2}H^{2}}\Gamma[t,{\bm p};Q,Q] 
                +\frac{B_{1}}{a^{2}H^{2}}\Gamma[t,{\bm p};Q,\Psi] 
                    \notag \\
    &\hspace{12em}%~~~~~~~~~~~~~~~~~~~~~~~~~~~~~~~~~~~~~~
        +\frac{B_{2}}{a^{2}H^{2}}\Gamma[t,{\bm p};Q,\Phi]
            \notag\\
    &\hspace{12em}%~~~~~~~~~~~~~~~~~~~~~~~~~~~~~~~~~~~~~~
        -\frac{M^{2}\alpha_{H}}{a^{2}H^{2}}\frac{1}{(2\pi)^{3}} \int d^{3}k_{1}d^{3}k_{2}\, 
            \delta^{(3)}({\bm k}_{1} +{\bm k}_{2}-{\bm p})k_{1}^{2}k_{2}^{2}\alpha({\bm k}_{1},{\bm k}_{2}) 
                Q(t,{\bm k}_{1})\Psi(t,{\bm k}_{2})
                    \notag\\
    &\hspace{12em}%~~~~~~~~~~~~~~~~~~~~~~~~~~~~~~~~~~~~~~
        +\frac{C_{0}}{a^{4}H^{4}}\frac{1}{(2\pi)^{6}}\int d^{3}k_{1}d^{3}k_{2}d^{3}k_{3}\, 
            \delta^{(3)}({\bm k}_{1}+{\bm k}_{2}+{\bm k}_{3}-{\bm p})
                \notag \\
    &\hspace{12em}%~~~~~~~~~~~~~~~~~~~~~~~~~~~~~~~~~~~~~~
    \times
        [ -k^{2}_{1}k^{2}_{2}k^{2}_{3} +3k^{2}_{1}({\bm k}_{2} \cdot {\bm k}_{3})^{2} 
            -2({\bm k}_{1} \cdot {\bm k}_{2})({\bm k}_{2} \cdot {\bm k}_{3})({\bm k}_{3} \cdot {\bm k}_{1}) ]
                Q(t,{\bm k}_{1})Q(t,{\bm k}_{2})Q(t,{\bm k}_{3}),
                    \label{Q F}
\end{align}
where for $Y,Z=\Psi, \Phi, Q$ we defined
\begin{align}
    \Gamma[t,{\bm p};Y,Z]  
        &= \frac{1}{(2\pi)^{3}}\int d^{3}k_{1}d^{3}k_{2} \delta^{(3)}({\bm k}_{1}+{\bm k}_{2}-{\bm p})  
            k_{1}^{2}k_{2}^{2}\gamma({\bm k}_{1}\cdot{\bm k}_{2}) Y(t,{\bm k}_{1})Z(t,{\bm k}_{2}),
\end{align}
\end{widetext}
and we introduced
\begin{align}
    \alpha({\bm k}_{1},{\bm k}_{2}) 
        & =1+\frac{({\bm k}_{1}\cdot{\bm k}_{2})}{k_{2}^{2}},
            \\
    \beta({\bm k}_{1},{\bm k}_{2})  
        & = \frac{({\bm k}_{1}\cdot{\bm k}_{2})|{\bm k}_{1}+{\bm k}_{2}|^{2}}{2k_{1}^{2}k_{2}^{2}}, 
            \\
    \gamma({\bm k}_{1},{\bm k}_{2}) 
        & = 1 -\frac{({\bm k}_{1}\cdot{\bm k}_{2})^{2}}{k_{1}^{2}k_{2}^{2}}.
\end{align}
The coefficients ${\cal F}_T,{\cal G}_T, A_1, A_2, \cdots$ all have the dimension of (mass)$^2$ and are written in terms of $M^2$ and the $\alpha$ parameters as presented explicitly in Appendix~\ref{sec: alphabets}.
One finds that there are four terms proportional to $\alpha_H$ in Eqs. (\ref{Psi F})--(\ref{Q F}) (the fourth term in the left hand side of Eq.~(\ref{Psi F}), the second term in the right hand side of Eq.~(\ref{Psi F}), the fourth term in the left hand side of Eq.~(\ref{Q F}), and the fourth term in the right hand side of Eq.~(\ref{Q F})).
Those are the new terms beyond Horndeski.
The other coefficients contain $\alpha_H$, but they are not new in the sense that even in the case of $\alpha_H=0$ those coefficients do not vanish and just reduce to the known expressions in the Horndeski theory~\cite{Takushima:2013foa}.
\footnotetext[3]{Our convention for the Fourier transform is
\begin{align}
    f(t,{\bm x}) 
        & = \frac{1}{(2\pi)^{3}}\int d^{3}p\ f(t,{\bm p})e^{i{\bm p}\cdot{\bm x}}.
            \notag
\end{align}}

\subsection{Fluid equations}

Since it is assumed that matter is minimally coupled to gravity,
the fluid equations are the same as the usual ones.
Under the quasi-static approximation, the conservation and Euler equations for nonrelativistic matter expressed in terms of the density contrast $\delta$
and the velocity field $u^{i}$ are given by
\begin{align}
    \dot{\delta} + \frac{1}{a}\nabla_{i}[(1+\delta)u^{i} ] 
        &= 0, 
            \label{continuous} \\
    \dot{u}^{i} + Hu^{i} +\frac{1}{a}u^{j}\nabla_{j}u^{i} 
        &=  -\frac{1}{a}\nabla^{i}\Phi. 
            \label{Eular}
\end{align}
Modification of gravity comes into play in the evolution of matter density perturbations
through the gravitational potential $\Phi$ in Eq.~(\ref{Eular}),
which is determined by Eqs.~(\ref{Psi F}), (\ref{Phi F}), and (\ref{Q F}).
Going to Fourier space, Eqs.~(\ref{continuous}) and (\ref{Eular}) are written as
\begin{align}
     &\frac{\dot{\delta}(t,{\bm p})}{H} +\theta(t,{\bm p}) 
         =  -\frac{1}{(2\pi)^{3}}\int d^{3}k_{1}d^{3}k_{2}\, 
            \delta^{(3)}({\bm k}_{1}+{\bm k}_{2}-{\bm p}) 
                \notag\\
        &\hspace{7em}
            \quad \times\alpha({\bm k}_{1},{\bm k}_{2})\theta(t,{\bm k}_{1})\delta(t,{\bm k}_{2}),
                \label{continuous F}  \\
    &\frac{\dot{\theta}(t,{\bm p})}{H} +\biggl(2+\frac{\dot H}{H^{2}} 
         \biggr)\theta(t,{\bm p}) -\frac{p^{2}}{a^{2}H^{2}}\Phi (t,{\bm p}) 
            \notag\\
         &\hspace{6.5em}~
            = -\frac{1}{(2\pi)^{3}}\int d^{3}k_{1}d^{3}k_{2}\, \delta^{(3)}({\bm k}_{1}+{\bm k}_{2}-{\bm p}) 
                \notag\\
        &\hspace{7.5em}~\,
            \times\beta({\bm k}_{1},{\bm k}_{2})\theta(t,{\bm k}_{1})\theta(t,{\bm k}_{2}), 
                \label{Eular F}
\end{align}
where we introduced a scalar function defined as $\theta = \nabla_{i} {u^{i}}/aH$.

\section{Matter density perturbations in GLPV theory}

Based on the set of the equations obtained in the previous section,
here, we derive the bispectrum of the matter density perturbations, $\delta$,
and highlight the impact of the new operators in the GLPV theory.
In order to investigate the matter bispectrum at the tree level,
we need to consider the perturbations up to second order under the assumption that the perturbations initially obey Gaussian statistics.
Before deriving the matter bispectrum from the second-order perturbations,
let us begin with giving a linear evolution equation for the matter density perturbations.

\subsection{Linear perturbations}
\label{sec:lin}

As we have shown in the previous section,
since we are considering the minimally-coupled matter there is not any modification in the continuity and Euler equations even in modified theories of gravity such as the GLPV theory.
Thus, the linear evolution equation for the matter density perturbations in Fourier space is
given by the standard one as
\begin{equation}
    \ddot{\delta} (t, {\bm p}) + 2 H \dot{\delta} (t, {\bm p}) + {p^2 \over a^2} \Phi = 0~.
\label{eq:lindelta}
\end{equation}
In the above equation,
modification of gravity is encoded in $\Phi$ that is determined from the modified Poisson equation.

Truncating Eqs.~(\ref{Psi F}),~(\ref{Phi F}), and (\ref{Q F}) at the linear order and solving them for $\Phi$, $\Psi$, and $Q$, one obtains the modified Poisson equation. However, even under the quasi-static approximation, those equations contain $\dot\Psi$ and $\dot Q$, and therefore it might not be straightforward to express $\Phi$ (and the other two variables) in terms of $\delta$. 
This can be done as follows: first,
Eqs.~(\ref{Psi F}) and~(\ref{Phi F}) (truncated at linear order) can be solved for $\Phi$ and $\Psi$;
then substituting the result to Eq.~(\ref{Q F}) (truncated at linear order) one obtains the equation written solely in terms of $Q$, $\delta$, and $\dot\delta$; in this equation, $\dot Q$'s are canceled and hence $Q$ can be expressed in terms of $\delta$ and $\dot \delta$; finally, using this result one can express $\Psi$ and $\Phi$ in terms of $\delta$, $\dot\delta$, and $\ddot\delta$.
The reason why this procedure works and in particular an algebraic equation is obtained for $Q$ is that the scalar-tensor theory we are considering is degenerate. 
The final result one thus arrives at is:
\begin{align}
    -\frac{p^2}{a^2H^2}Q    
        & = \kappa_Q\delta+\nu_Q\frac{\dot\delta}{H},
            \label{eq:Q=delta} \\
    -\frac{p^2}{a^2H^2}\Psi 
        & = \kappa_\Psi\delta+\nu_\Psi\frac{\dot\delta}{H},
            \label{eq:Psi=delta} \\
    -\frac{p^2}{a^2H^2}\Phi 
        & = \kappa_\Phi\delta+\nu_\Phi\frac{\dot\delta}{H}+\mu_\Phi\frac{\ddot\delta}{H^2},
            \label{eq:Phi=delta}
\end{align}
where the coefficients are given by
\begin{align}
    \nu_Q 
        & =  \frac{3}{2} M^2 \Omega_{\rm m} \, \frac{M^2 \alpha_H \mathcal{G}_T}{\mathcal{Z}},
            \\
    \kappa_Q 
        & =  \frac{3}{2} M^2 \Omega_{\rm m} \, \frac{\mathcal{T}}{\mathcal{Z}},
            \\
    \nu_\Psi 
        & = -\frac{3}{2} M^2 \Omega_{\rm m} \, \frac{M^2 \alpha_H A_2}{\mathcal{Z}}, 
            \\
    \kappa_\Psi 
        & = \frac{3}{2} M^2 \Omega_{\rm m} \, \frac{\mathcal{S}}{\mathcal{Z}}, 
            \\
    \mu_\Phi 
        & = \frac{M^2 \alpha_H}{\mathcal{G}_T} \nu_Q, 
            \\
    \nu_\Phi 
        & = \frac{1}{\mathcal{G}_T} \biggl\{ \mathcal{F}_T \nu_\Psi - A_3 \nu_Q 
            \nonumber\\ 
        & \hspace{3.5em} + M^2 \alpha_H \left[ \kappa_Q + 
            \frac{1}{a^2 H^2} \left( a^2 H \nu_Q \right)^\dd \right] \biggr\},
                \\
    \kappa_\Phi 
        & = \frac{1}{\mathcal{G}_T} \left\{ \mathcal{F}_T \kappa_\Psi - A_3 \kappa_Q
            + \frac{M^2 \alpha_H}{a^2 H^3} \left( a^2 H^2 \kappa_Q \right)^\dd \right\}.
\end{align}
with
\begin{align}
    &\mathcal{T} 
        := A_2 \,\mathcal{F}_T + A_1 \, \mathcal{G}_T 
            - M^2 \alpha_H \left( \mathcal{G}_T + \frac{\dot{\mathcal{G}}_T}{H} \right), 
                \\
    & \mathcal{S} 
        := A_0 \,\mathcal{G}_T + A_2 \,A_3 + M^2 \alpha_H \left(A_2 + \frac{\dot{A}_2}{H} \right),  
            \\
    & \mathcal{Z} 
        := A_0\,\mathcal{G}_T^2 + A_2 (A_1 + A_3) \mathcal{G}_T + A_2^2 \,\mathcal{F}_T
            \notag \\ 
    &\hspace{2em}
        + \frac{M^2 \alpha_H}{H}{\cal G}_{T}^{2}\left( \frac{A_2}{\mathcal{G}_T} \right)^\dd .
\end{align}
Equation~(\ref{eq:Phi=delta}) allows us to eliminate $\Phi$ from Eq.~(\ref{eq:lindelta}), 
leaving a closed-form, second-order evolution equation for $\delta$:
\begin{align}
\ddot{\delta} + (2+\varsigma)H\dot{\delta} -\frac{3}{2}\Omega_{\mathrm{m}}\Xi_{\Phi}H^{2}\delta = 0\label{first-order eq.},
\end{align}
where $\varsigma(t)$ and $\Xi_{\Phi}(t)$ are defined by
\begin{align}
\varsigma & :=\frac{2 \mu_\Phi - \nu_\Phi}{1 - \mu_\Phi}, \\
\frac{3}{2} \Omega_{\rm m} \, \Xi_\Phi  &:= \frac{\kappa_\Phi}{1 - \mu_\Phi} .
\end{align}
In the Horndeski limit, $\alpha_H=0$, one finds that the additional friction term vanishes, $\varsigma=0$.
It is easy to confirm that in the same limit $\Xi_\Phi$ reproduces the previous result~\cite{Kobayashi:2014ida,DAmico:2016ntq}.\footnote{Our notation $(\varsigma,\Xi_{\Phi})$ translates to $(\gamma,\mu_{\Phi})$ in Ref.~\cite{DAmico:2016ntq}.}
We write the growing solution to Eq.~(\ref{first-order eq.}) as
\begin{align}
\delta(t, {\bf p}) = D_+(t)\delta_{\rm L}({\bf p}),
\end{align}
where $\delta_{\rm L}({\bf p})$ represents the initial linear density field.
The effect of modification of gravity is thus separated and imprinted in the evolution of the matter density perturbations, $D_+(t)$.
We also introduce the linear growth rate, $f:=d\ln D_+/d\ln a$, which is often used in the literature.

Using Eq.~(\ref{first-order eq.}), one can eliminate $\ddot\delta$ from Eq.~(\ref{eq:Phi=delta}). 
Then, replacing $\dot \delta$ with $fH\delta$, we can rewrite Eqs.~(\ref{eq:Q=delta})--(\ref{eq:Phi=delta}) as
\begin{align}
-\frac{p^2}{a^2H^2}Q & = \left( \kappa_Q + f \nu_Q \right) \delta =: K_Q \,\delta,\label{eq:QKQ} \\
-\frac{p^2}{a^2H^2}\Psi & = \left( \kappa_\Psi + f \nu_\Psi \right)\delta =: K_\Psi \, \delta, \label{Phi_F_1}\\
-\frac{p^2}{a^2H^2}\Phi & = \left( \frac{3}{2}\Omega_{\rm m}\Xi_{\Phi} -\varsigma f \right) \delta =: K_\Phi \, \delta. \label{Psi_F_1}
\end{align}
These equations are convenient for the second-order analysis in the next subsection.

\subsection{Second-order perturbations}

To investigate the bispectrum of $\delta$ at the tree level, we need to solve the perturbation equations up to second order.
Let us now move to the second-order analysis of the matter density perturbations based on
the equations derived in the previous section.
Substituting the first-order solutions~(\ref{eq:QKQ})--(\ref{Psi_F_1}) to the right hand sides of
Eqs.~(\ref{Psi F})--(\ref{Q F}), we obtain, up to second order in $\delta$,
\begin{widetext}
\begin{align}
{\cal F}_{T}\Psi -{\cal G}_{T}\Phi -A_{3}Q +M^2\alpha_H\frac{{\dot Q}}{H}
&= -D_{+}^{2}\frac{a^{2}H^{2}}{p^{2}} \left( M^2\alpha_HK_{Q}^{2}\,{\cal W}_{\beta}({\bm p}) + \frac{B_{1}}{2} K_{Q}^{2}\,{\cal W}_{\gamma} ({\bm p}) \right), \label{Psi F_{2}}\\
{\cal G}_{T}\Psi+A_{2}Q  +\frac{a^{2}}{2p^{2}}\rho_{\rm m}\delta
&=  D_{+}^{2}\frac{a^{2}H^{2}}{p^{2}}\frac{B_{2}}{2}K_{Q}^{2}\,{\cal W}_{\gamma}({\bm p}),  \label{Phi F_{2}}\\
A_{0}Q -A_{1}\Psi -A_{2}\Phi -M^2\alpha_H\frac{\dot \Psi}{H}
&=  D_{+}^{2}\frac{a^{2}H^{2}}{p^{2}}\left[ M^2\alpha_HK_{Q}K_{\Psi}\,{\cal W}_{\alpha} ({\bm p})
+ \left(B_{0}K_{Q}^{2} -B_{1}K_{\Psi}K_{Q} -B_{2}K_{\Phi}K_{Q}\right){\cal W}_{\gamma} ({\bm p}) \right],\label{Q F_{2}}
\end{align}
where
${\cal W}_\alpha({\bm p}):={\cal I}[{\bm p};\alpha_s({\bm k}_1,{\bm k}_2)]$,
${\cal W}_\beta({\bm p}):={\cal I}[{\bm p};\beta({\bm k}_1,{\bm k}_2)]$, and
${\cal W}_\gamma({\bm p}):={\cal I}[{\bm p};\gamma({\bm k}_1,{\bm k}_2)]$,
with
\begin{align}
{\cal I}[{\bm p};Y({\bm k}_1,{\bm k}_2)]:=
\frac{1}{(2\pi)^{3}}\int d^{3}k_{1}d^{3}k_{2}\, \delta^{(3)}({\bm k}_{1}+{\bm k}_{2}-{\bm p}) \,
Y({\bm k}_{1},{\bm k}_{2})\delta_{\rm L}({\bm k_{1}})\delta_{\rm L}({\bm k_{2}}).
\end{align}
\end{widetext}
Here we introduced a symmetrized version of $\alpha ({\bm k}_1, {\bm k}_2)$ as
\begin{align}
\alpha_{s}({\bm k}_{1},{\bm k}_{2}) & = 1+\frac{({\bm k}_{1}\cdot{\bm k}_{2})(k_{1}^{2}+k_{2}^{2})}{2k_{1}^{2}k_{2}^{2}}.
\end{align}
Note that we have the following relation:
${\cal W}_{\beta}({\bm p})={\cal W}_{\alpha}({\bm p})-{\cal W}_{\gamma}({\bm p})$.
The functions ${\cal W}_\alpha, {\cal W}_\beta$, and ${\cal W}_\gamma$ are dependent on the initial density field $\delta_{\rm L}({\bm k})$, but not on modification of gravity.

From the nonlinear fluid equations (\ref{continuous F}) and (\ref{Eular F}) with the analysis of the linear perturbations in \ref{sec:lin},
we can obtain the following equation up to the second order in $\delta_{\rm L}$:
\begin{align}
\ddot{\delta} + 2 H \dot{\delta} + \frac{p^2}{a^2} \Phi = H^2 D_+^2 \left( \tilde{S}_\alpha {\cal W}_\alpha - \tilde{S}_\gamma {\cal W}_\gamma \right),
\label{second_1}
\end{align}
where
\begin{align}
&\tilde{S}_\alpha = 2 f^2 + \frac{3}{2} \Omega_{\rm m} \Xi_\Phi - \varsigma \, f , \\
&\tilde{S}_\gamma = f^2.
\end{align}
The second-order nonlinearity due to the modification of gravity, which appears in the right hand sides of Eqs.~(\ref{Psi F_{2}})--(\ref{Q F_{2}}), is introduced through the gravitational potential $\Phi$ as follows.
Repeating the same procedure as in the linear analysis, we can write $Q$, $\Phi$, and $\Psi$ in terms of $\delta$, its first and second derivatives,
and the second-order terms in the right hand sides of Eqs.~(\ref{Psi F_{2}})--(\ref{Q F_{2}}) as
\begin{align}
-\frac{p^2}{a^2H^2}Q & =\kappa_Q\delta+\nu_Q\frac{\dot\delta}{H}+D_+^2\left(\tau_{Q\alpha} {\cal W}_\alpha - \tau_{Q\gamma} {\cal W}_\gamma \right),\label{eq:secondQ} \\
-\frac{p^2}{a^2H^2}\Psi & = \kappa_\Psi\delta+\nu_\Psi\frac{\dot\delta}{H}+ D_+^2 \left( \tau_{\Psi\alpha} {\cal W}_\alpha - \tau_{\Psi\gamma} {\cal W}_\gamma \right),\label{eq:secondPsi} \\
-\frac{p^2}{a^2H^2}\Phi & = \kappa_\Phi\delta+\nu_\Phi\frac{\dot\delta}{H}+\mu_\Phi\frac{\ddot\delta}{H^2} \notag \\ 
&\quad +D_+^2 \left(\tau_{\Phi\alpha} {\cal W}_\alpha - \tau_{\Phi\gamma} {\cal W}_\gamma \right), \label{eq:secondPhi}
\end{align}
where
\begin{widetext}
\begin{align}
    \tau_{Q \alpha} 
        &= - \frac{M^2 \alpha_H}{\mathcal Z} 
            \left( A_2 {\mathcal G}_T K_Q^2 + {\mathcal G}_T^2 K_Q K_\Psi \right),    
                \\
    \tau_{Q \gamma} 
        &=  \frac{1}{\mathcal Z} \left\{ \left[ B_0 {\mathcal G}_T^2 + \frac{B_1}{2} A_2 {\mathcal G}_T  
            + \frac{B_2}{2} \left({\mathcal T} + 3 M^2 \alpha_H{\mathcal G}_T 
                \left( 1 + \frac{2}{3}\frac{\dot{H}}{H^2} \right)\right) - M^2 \alpha_H A_2 {\mathcal G}_T \right] 
                    K_Q^2 \right.
                        \nonumber\\
        & \left.\hspace{3.5em} - B_1 {\mathcal G}_T^2  K_\Psi K_Q - B_2 {\mathcal G}_T^2 K_\Phi K_Q 
            + \frac{M^2 \alpha_H {\mathcal G}_T}{2} \frac{(D_+^2 B_2 K_Q^2)^\dd}{D_+^2 H}  \right\}, 
                \\
    \tau_{\Psi \alpha} 
        &=  \frac{M^2 \alpha_H}{\mathcal Z}  \left( A_2^2 K_Q^2 + A_2{\mathcal G}_T K_Q K_\Psi \right),
            \\
    \tau_{\Psi \gamma} 
        &=\frac{B_{2}K_{Q}^{2} -2A_{2}\tau_{Q\gamma}}{2{\cal G}_{T}} 
            \notag\\
        &= -\frac{1}{\mathcal Z} \left\{ \left[B_0 A_2 {\mathcal G}_T + \frac{B_1}{2} A_2^2 + \frac{B_2}{2} 
            \left(-{\mathcal S} + 3 M^2 \alpha_H A_2 \left( 1 + \frac{2}{3}\frac{\dot{H}}{H^2} \right)\right) 
                - M^2 \alpha_H A_2^2 \right] K_Q^2 \right.
                    \nonumber\\
        & \left.\hspace{3.5em} - B_1 A_2 {\mathcal G}_T  K_\Psi K_Q - B_2 A_2 {\mathcal G}_T K_\Phi K_Q 
            + \frac{M^2 \alpha_H A_2}{2} \frac{(D_+^2 B_2 K_Q^2)^\dd}{D_+^2 H} \right\}, 
                \\
    \tau_{\Phi \alpha} 
        &= \frac{1}{{\mathcal G}_T}\left\{{\mathcal F}_T \, \tau_{\Psi \alpha} 
            - \left[A_3 - 2 M^2 \alpha_H \left( 1 + f+\frac{\dot{H}}{H^2} \right) \right] \, \tau_{Q \alpha} 
                + M^2 \alpha_H \frac{\dot{\tau}_{Q \alpha}}{H} - M^2 \alpha_H K_Q^2\right\},
                    \\
    \tau_{\Phi \gamma} 
        &=  \frac{1}{{\mathcal G}_T} \left\{{\mathcal F}_T \, \tau_{\Psi \gamma}
            - \left[A_3 - 2 M^2 \alpha_H \left( 1 + f+\frac{\dot{H}}{H^2} \right) \right] \, 
                \tau_{Q \gamma} + M^2 \alpha_H \frac{\dot{\tau}_{Q \gamma}}{H} -\left( \frac{B_1}{2} 
                    -M^2 \alpha_H \right) K_Q^2 \right\}.
\end{align}
We then eliminate $\Phi$ from Eq. (\ref{second_1}) and obtain the evolution equation for $\delta$ capturing the effect of the second-order nonlinearity of the scalar field:
\begin{align}
    \ddot\delta + (2+\varsigma)H\dot\delta -\frac{3}{2}\Omega_{\mathrm{m}}\Xi_{\Phi}H^{2}\delta 
        = D_{+}^{2}H^{2} \left(S_{\alpha}{\cal W}_{\alpha} - S_{\gamma} {\cal W}_{\gamma} \right)
            \label{second-order eq.}.
\end{align}
In the right hand side we defined $S_\alpha$ and $S_\gamma$ by
\begin{align}
    \left( 1 - \mu_\Phi \right)S_{\alpha}(t) 
        & :=  \tilde{S}_\alpha +\tau_{\Phi\alpha},
            \label{source1} \\
    \left( 1 - \mu_\Phi \right)S_\gamma (t) 
        & :=  \tilde{S}_\gamma +\tau_{\Phi\gamma}, 
            \label{source2}
\end{align}
The second-order nonlinearity due to modification of gravity appears in all of these $\tau$ coefficients, but it should be emphasized that $\tau_{\Phi\alpha}= 0$ for $\alpha_H= 0$ ({\em i.e.}, in the Horndeski theory ), while $\tau_{\Phi\gamma }\neq 0$ in general if gravity is modified anyway (see Table~\ref{tab:}). 
In other words, $\tau_{\Phi\alpha}$ is a new term beyond Horndeski.
The solution to Eq.~(\ref{second-order eq.}) up to second order in $\delta_{\rm L}$ can be written as
\begin{align}
    \delta(t,{\bm p}) 
        = D_+(t) \delta_{\rm L}({\bm p}) + D_{+}^{2}(t) \frac{1}{(2\pi)^{3}}\int d^{3}k_{1}d^{3}k_{2}\, 
            \delta^{(3)}({\bm k}_{1}+{\bm k}_{2}-{\bm p}) F_2 (t,{\bm k}_1,{\bm k}_2) 
                \delta_{\rm L}({\bm k_{1}})\delta_{\rm L}({\bm k_{2}}),
                    \label{delta_{2}}
\end{align}
with the second-order kernel defined as
\begin{align}
    F_{2}(t,{\bm k}_{1},{\bm k}_{2}) 
        := \kappa(t)\alpha_{s}({\bm k}_{1},{\bm k}_{2}) - \frac{2}{7}\lambda(t)\gamma({\bm k}_{1},{\bm k}_{2}),         
            \label{kernel}
\end{align}
where $\kappa (t)$ and $\lambda (t) $ are the solutions of the following second-order differential equations,
\begin{align}
    & \ddot\kappa + [4f +(2 + \varsigma) ]H\dot\kappa 
        +H^{2}\left(2f^{2}+\frac{3}{2}\Omega_{\mathrm{m}}\Xi_{\Phi}\right)\kappa 
            = H^{2}S_\alpha, 
                \label{eq:derikappa}\\
    & \ddot\lambda + [4f +(2+\varsigma)]H\dot\lambda 
        +H^{2}\left(2f^{2}+\frac{3}{2}\Omega_{\mathrm{m}}\Xi_{\Phi}\right)\lambda 
            = \frac{7}{2} H^{2}S_\gamma, 
                \label{eq:derilambda}
\end{align}
\end{widetext}
supplemented with the condition that $\kappa, \lambda \to 1$ in the early time (it is easy to check that $\kappa=\lambda = 1$ indeed solves Eqs.~(\ref{eq:derikappa})  and~(\ref{eq:derilambda}) if all the $\alpha$ parameters
are negligibly small and $\Omega_{\rm m}=1$).
In the Horndeski limit ($\alpha_H = 0$), these expressions reproduce the result of Ref.~\cite{Takushima:2013foa}.
Especially, since $\varsigma = 0$ and $\tau_{\Phi \alpha} = 0$ in the Horndeski theory (see Table~\ref{tab:}),
the right hand side of Eq.~(\ref{eq:derikappa}) reduces to $H^2\left(2f^{2} +3\Omega_{\mathrm{m}}\Xi_{\Phi}/2\right)$, so that $\kappa(t)=1$ at any time.
In this case the second-order kernel~(\ref{kernel}) therefore depends only on $\lambda(t)$~\cite{Takushima:2013foa}.
Thus, we find that a new feature in the GLPV theory beyond Horndeski is the $\kappa$ term that is different from 1 and is time-dependent in general.
This is the main result of this paper.

%%%%%%%%%%%%%%%%%%
\begin{table}[htb]
  \begin{tabular}{|c|c|c|c|} \hline
     & $\Lambda$CDM & Horndeski & beyond   \\ \hline \hline
     $\varsigma$ & 0 & 0 & \checkmark   \\ \hline
     $\Xi_\Phi$ & 1 & \checkmark & \checkmark   \\ \hline
     $\mu_\Phi$ & 0 & 0 & \checkmark  \\ \hline
     $\tau_{\Phi \alpha}$ & 0 & 0 & \checkmark  \\ \hline
     $\tau_{\Phi \gamma}$ & 0 & \checkmark & \checkmark   \\ \hline
  \end{tabular}
  \caption{Summary of the parameters in the second-order evolution equation for $\delta$, (\ref{second-order eq.}) with Eqs. (\ref{source1}) and (\ref{source2}).}
  \label{tab:}
\end{table}
%%%%%%%%%%%%%%%%%%

\section{Matter bispectrum}

%%%%%%%%%%%%%%%%%%%%%%%%%%%%%%%%%%%%%%%%%%%%%%%%%
\begin{figure*}[htp]
    \centering
        \subfigure{
            \includegraphics[width = 7.5cm]{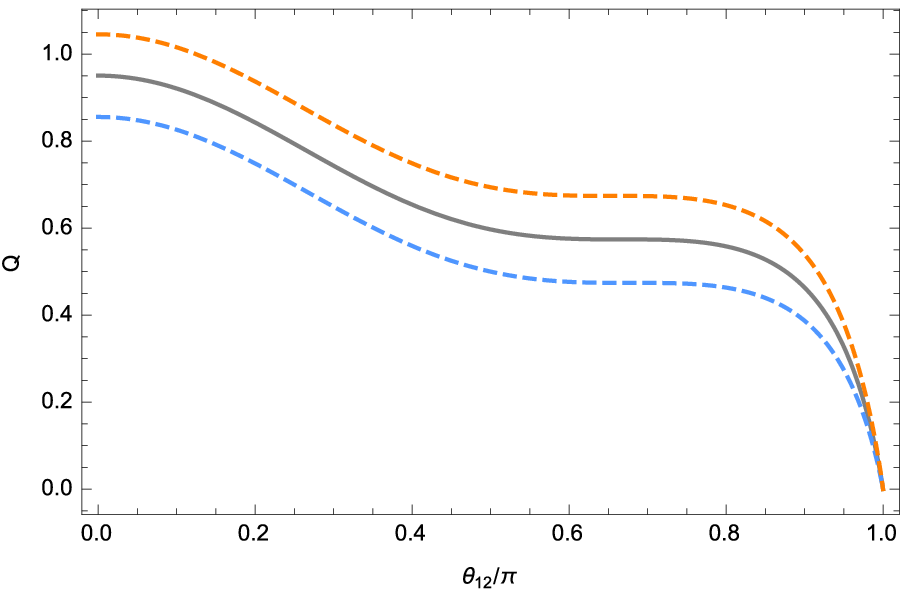}
                \label{glpv(a)}
            }
        \subfigure{
            \includegraphics[width = 7.5cm]{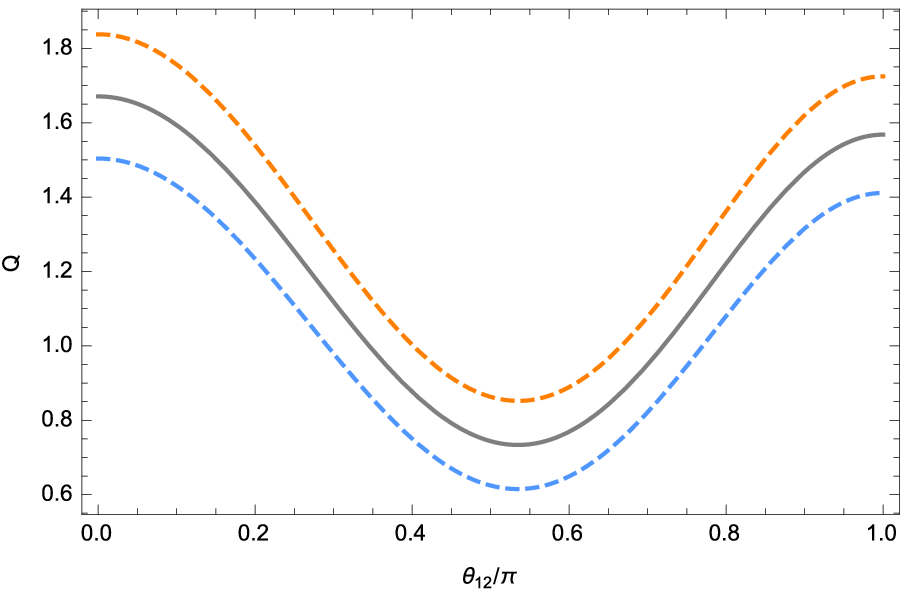}
                \label{glpv(b)}
            }
        \vspace{1cm}
    \caption{(color online) The reduced bispectrum as a function of $\theta_{12}$, with fixed $k_1$ and $k_2$.
    We adopt the isosceles triangular configuration with $k_{1}=k_{2}=0.01h/{\rm Mpc}$ in the left panel (a),
    and the distorted triangle with $k_{1}=5 k_{2}=0.05h/{\rm Mpc}$ in the right panel (b).
    In both panels, we take a different value for $\kappa (t)$ to be $1.0$ (gray solid line), $0.9$ (blue dashed line), 
    and $1.1$ (orange dashed line), while $\lambda (t)$ is fixed to be 1.}%
\end{figure*}
%%%%%%%%%%%%%%%%%%%%%%%%%%%%%%%%%%%%%%%%%%%%%%%%%

%%%%%%%%%%%%%%%%%%%%%%%%%%%%%%%%%%%%%%%%%%%%%%%%%
\begin{figure*}[htp]
    \centering
        \subfigure{
            \includegraphics[width = 7.5cm]{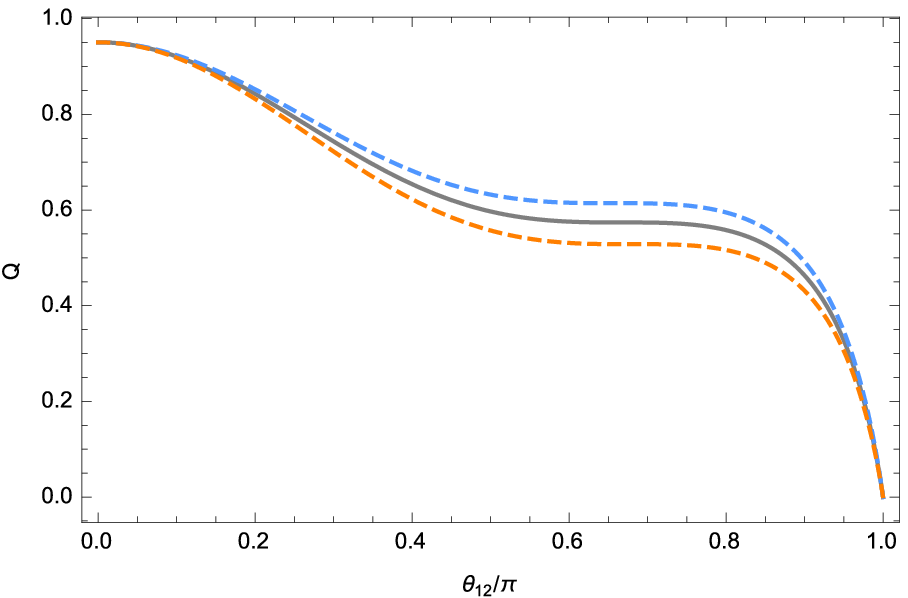}
                \label{horn(a)}
            }
        \subfigure{
            \includegraphics[width = 7.5cm]{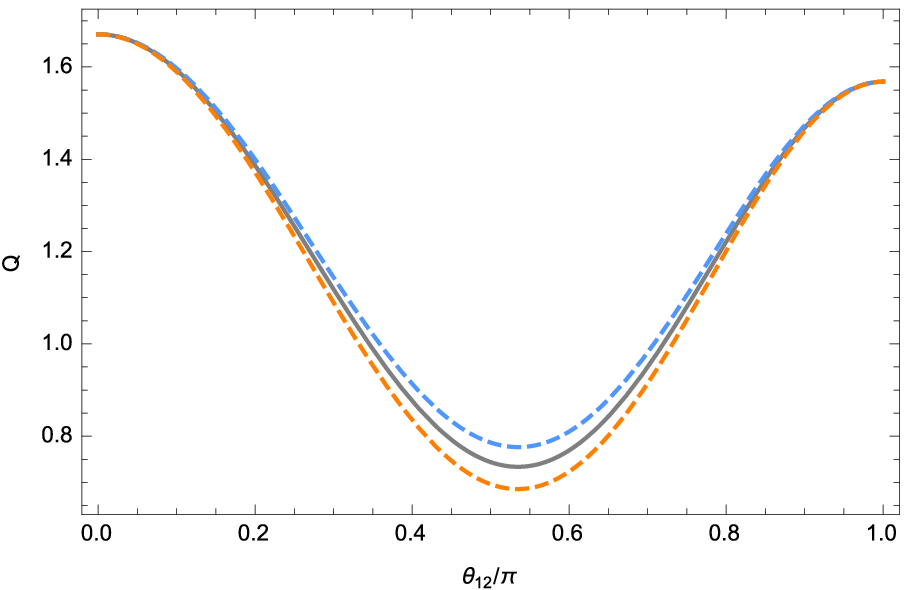}
                \label{horn(b)}
            }
        \vspace{1cm}
    \caption{(color online) The reduced bispectrum as a function of $\theta_{12}$, with fixed $k_1$ and $k_2$.
    We adopt the isosceles triangular configuration with $k_{1}=k_{2}=0.01h/{\rm Mpc}$ in the left panel (a),
    and the distorted triangle with $k_{1}=5 k_{2}=0.05h/{\rm Mpc}$ in the right panel (b).
    In both panels, we take a different value for $\lambda (t)$ to be $1.0$ (gray solid line), $0.9$ (blue dashed line), 
    and $1.1$ (orange dashed line), while $\kappa (t)$ is fixed to be 1.}%
\end{figure*}
%%%%%%%%%%%%%%%%%%%%%%%%%%%%%%%%%%%%%%%%%%%%%%%%%

Finally,
let us investigate the matter bispectrum as an observable for probing such quasi-nonlinear evolution based on the above analysis for the matter density perturbations up to second order.
The power spectrum and the bispectrum of the matter density perturbations are respectively defined by
\begin{align}
    & \langle \delta(t,{\bm k}_{1})\delta(t,{\bm k}_{2}) \rangle 
        =: (2\pi)^{3}\delta^{(3)}({\bm k}_{1}+{\bm k}_{2})P(t,k_{1}), 
            \\
    & \langle \delta(t,{\bm k}_{1})\delta(t,{\bm k}_{2})\delta(t,{\bm k}_{3}) \rangle 
        =: (2\pi)^{3}\delta^{(3)}({\bm k}_{1}+{\bm k}_{2}+{\bm k}_{3}) 
            \notag\\
    &\hspace{11.5em} \times B(t,k_{1},k_{2},k_{3}).
\end{align}
Here, for simplicity we assume that the initial density field $\delta_{\rm L}$ obeys Gaussian statistics, and,
by making use of the expression (\ref{delta_{2}}), the matter bispectrum at the tree-level can be evaluated as
\begin{align}
\label{treebis}
    & D_{+}^{-4}(t)B(t,k_{1},k_{2},k_{3}) 
        \notag\\
    & =2[F_{2}(t,{\bm k}_{1},{\bm k}_{2})P_{11}(k_{1})P_{11}(k_{2}) +{\rm 2\ cyclic\ terms}],
\end{align}
where $P_{11}$ represents the power spectrum of the initial density field defined by
\begin{align}
    \langle \delta_{L}({\bm k}_{1})\delta_{L}({\bm k}_{2})\rangle 
        =: (2\pi)^{3}\delta^{(3)}({\bm k}_{1}+{\bm k}_{2})P_{11}(k_{1}).
\end{align}
As we have mentioned before, we assume that standard cosmology is recovered in the early stage of the matter-dominant universe.
Thus, here, we calculate $P_{11}(k)$ adopting the best fit cosmological parameters taken from Planck data~\cite{Ade:2015xua}.

As usual, in order to investigate the shape of the bispectrum in Fourier space, let us introduce
a reduced bispectrum which is defined by
\begin{align}
    & Q_{123}(t,k_{1},k_{2},k_{3}) 
        \notag\\
    & := \frac{B(t,k_{1},k_{2},k_{3})}{D^{4}_+(t)[P_{11}(k_{1})P_{11}(k_{2}) +{\rm 2\ cyclic\ terms}]}.
\end{align}
From Eq.~(\ref{treebis}), we have
\begin{align}
    & Q_{123}(t,k_{1},k_{2},k_{3}) 
        \notag \\
    & = \frac{2[F_{2}(t,{\bm k}_{1},{\bm k}_{2})P_{11}(k_{1})P_{11}(k_{2}) +{\rm 2\ cyclic\ terms}]}
        {[P_{11}(k_{1})P_{11}(k_{2}) +{\rm 2\ cyclic\ terms}]}.
\end{align}
Thus, the reduced bispectrum does not depend on the linear growth function $D_+$ and the effect of modification of gravity is encoded in the second-order kernel, $F_2(t, {\bm k}_1,{\bm k}_2)$.
As we have discussed, the characteristic feature of the GLPV theory beyond Horndeski with $\alpha_H \neq 0$ is that $\kappa$ in the second-order kernel is different from 1 and is time-dependent in general.

To demonstrate how this new feature beyond Horndeski distorts the shape of the bispectrum,
we plot in Figs.~\ref{glpv(a)} and \ref{glpv(b)} the reduced bispectrum as a function of $\theta_{12}$ which is the angle between ${\bm k}_1$ and ${\bm k}_2$, with $k_1$ and $k_2$ fixed.
In these figures, we take different values of $\kappa$ as $\kappa=1.0$ (gray solid line), $0.9$ (blue dashed line), and $1.1$ (orange dashed line) while we fix $\lambda=1.0$.
As one can see, except for the squeezed configurations ($\theta_{12} \to \pi$ in Fig.~\ref{glpv(a)}), the reduced bispectrum becomes larger for $\kappa > 1$ and smaller for $\kappa < 1$.

As a comparison,
we show in Figs.~\ref{horn(a)} and \ref{horn(b)} the reduced bispectrum for different values of $\lambda$.
In these figures, $\kappa$ is fixed to be 1 which corresponds to the case with $\alpha_H = 0$.
Compared with Figs.~\ref{glpv(a)} and~\ref{glpv(b)}, one finds that the deviation of $\lambda$ from unity would give a large effect on the reduced bispectrum only for $\theta_{12} \simeq 2 \pi / 3$ in the left panel and  $\theta_{12} \simeq \pi / 2$ in the right panel.
Thus, the effect of the GLPV theory on the matter bispectrum is significant for $\theta_{12} \to 0$.
In other words,
the matter bispectrum with $\theta_{12} = 0$ is considered to be a powerful probe of the GLPV theory beyond Horndeski.

\section{After GW170817}

The gravitational wave event GW170817~\cite{TheLIGOScientific:2017qsa}
and its optical counterpart GRB 170817A~\cite{Monitor:2017mdv} placed a tight constraint on the propagation speed of gravitational waves, $|c_T-c|<{\cal O}(10^{-15})$.
The consequences of this constraint on the Galileon theory, the Horndeski theory, and its extensions have been discussed in Refs.~\cite{Crisostomi:2017lbg,Langlois:2017dyl,Dima:2017pwp,Crisostomi:2017pjs,Creminelli:2017sry,Ezquiaga:2017ekz,Baker:2017hug,Sakstein:2017xjx,Arai:2017hxj,Peirone:2017vcq,Amendola:2017orw,Peirone:2017ywi,Kreisch:2017uet,Cusin:2017mzw,Bartolo:2017ibw,Amendola:2017ovw,Kase:2018iwp}.\footnote{See Refs.~\cite{Jimenez:2015bwa, Lombriser:2015sxa, Brax:2015dma, Lombriser:2016yzn, Bettoni:2016mij, Sawicki:2016klv} for earlier works before this event on
the prospects of measuring $c_T$}
In terms of the functions in the action, the constraint reads
\begin{align}
|\alpha_T|<{\cal O}(10^{-15})
\quad \Rightarrow \quad G_{4X}+XF_4\simeq 0.
\end{align}
This must hold at least in the late-time universe.
Upon imposing $\alpha_T=0$, we have $\alpha_H=\alpha_G$,
while $\alpha_M$, $\alpha_B$, and $\alpha_H$ itself are still allowed to be ${\cal O}(0.1) - {\cal O}(1)$~\cite{Sakstein:2017xjx}.
A further constraint can be obtained from the Hulse-Taylor pulsar under the additional assumption that the scalar radiation does not take part in the energy loss, which leads to $|\alpha_H|<{\cal O}(10^{-3})$~\cite{Dima:2017pwp}.
This implies that ${\cal O}(10^{-3})$--${\cal O}(1)$ deviation of $\kappa$ from its Horndeski value ($\kappa=1$) is still possible, depending on the assumption one makes.

\section{Conclusions}

In this paper, we have studied the matter bispectrum of large scale structure as a probe of the so-called ``beyond Horndeski'' theory or the GLPV theory of modified gravity.
We focused on the nonlinearity generated from derivative interactions of the metric perturbations and the scalar degree of freedom and derived a second-order solution of the matter density perturbations $\delta(t,{\bm k})$.
We have shown that a new, time-dependent coefficient $\kappa$ appears in the second-order kernel in the GLPV theory.
Since we have $\kappa := 1$ in general relativity and even in the Horndeski theory~\cite{Takushima:2013foa}, 
this is certainly a characteristic feature of the theory beyond Horndeski.
Based on this second-order solution, we have evaluated the matter bispectrum and found that the effect of nonstandard values of $\kappa$ can be seen in the bispectrum at the folded configurations ($k_1+k_2=k_3$).
We thus conclude that a deformed matter bispectrum at the folded configurations can be a unique probe of ``beyond Horndeski'' operators.
Note that there exist several scenarios where the primordial curvature perturbations would acquire the folded-type non-Gaussianity during inflation (see, {\it e.g.}, Ref.~\cite{Ade:2015ava}) and such a type of primordial non-Gaussianity could also deform the matter bispectrum at the folded configurations.
However, if we can precisely measure not only the dependence of $\theta_{12}$ but also the scale dependence of the matter bispectrum, it would help us discriminate the signature of beyond Horndeski from such a folded-type primordial non-Gaussianity.

In light of the recent gravitational wave event GW170817, 
there is a growing interest in the so-called DHOST theories which are more general than the one considered in this paper but evade the stringent constraint on the propagation speed of gravitational waves~\cite{Crisostomi:2017lbg, Langlois:2017dyl, Dima:2017pwp, Crisostomi:2017pjs}.
It would be interesting to explore how the results in the present work can be extended to such DHOST theories.

%%%%%%%%%%%%%%%%%%%%%%%%%%%%%%%%%%%%%%%%%%%%%%%%%
\acknowledgements
%%%%%%%%%%%%%%%%%%%%%%%%%%%%%%%%%%%%%%%%%%%%%%%%%

We would like to thank
Kazuyuki Akitsu, Takashi Hiramatsu, Rampei Kimura, Sakine Nishi, Shinji Mukohyama,
Masahiro Takada, Shinji Tsujikawa, and
Daisuke Yamauchi for useful comments and fruitful discussion.
This work was supported in part by the
JSPS Research Fellowships for Young Scientists
No.~17J04865 (S.H.),
the JSPS Grants-in-Aid for Scientific Research
 Nos.~16H01102, 16K17707 (T.K), 15K17659, 16H01103 (S.Y.), 15K17646, and 17H01110~(H.T.),
 MEXT-Supported Program for the Strategic Research Foundation at Private Universities,
 2014-2017 (S1411024) (T.K. and S.Y.),
and MEXT KAKENHI Grant Nos.~15H05888 (T.K. and S.Y.) and~17H06359 (T.K.).

%%%%%%%%%%%%%%%%%%%%%%%%%%%%%%%%%%%%%%%%%%%%%%%%%
\appendix
%%%%%%%%%%%%%%%%%%%%%%%%%%%%%%%%%%%%%%%%%%%%%%%%%

\section{The sound speed of the scalar mode and the limit of the quasi-static approximation}
\label{app: cs2}

When matter is present, the sound speed of the scalar field is given by~\cite{Gleyzes:2014qga}
\begin{align}
c_s^2 &= \frac{2(1+\alpha_{B})^{2}}{\alpha_K +6\alpha_B^2} \Biggl[ \alpha_{M} -\alpha_{T} -\frac{\dot H}{H^{2}} +\frac{1}{H}\frac{d}{dt}\biggl(\frac{1+\alpha_{H}}{1+\alpha_{B}}\biggr)  
 \notag\\
&\quad +\frac{\alpha_{H}-\alpha_{B}}{1+\alpha_{B}} \biggl(\alpha_{M}+1-\frac{\dot H}{H^{2}}\biggr) -\frac{3\Omega_{\rm m}}{2}\left(\frac{1+\alpha_{H}}{1+\alpha_{B}}\right)^{2}\ \Biggr],
\end{align}
where we defined
\begin{align}
H^{2}M^{2}\alpha_{K} & =
2X(G_{2X} +2XG_{2XX} -2G_{3\phi} -2XG_{3\phi X}) \notag\\
&\quad +12HX\dot\phi (G_{3X} +XG_{3XX} -3G_{4\phi X} \notag \\
& \quad -2XG_{4\phi XX})\notag \\
& \quad +12H^{2}X (G_{4X} +8XG_{4XX} +4X^{2}G_{4XXX}) \notag \\
& \quad +24H^{2}X^{2} (6F_{4} +9XF_{4X} +2X^{2}F_{4XX}).
\end{align}
In this paper we assume that the quasi-static approximation is valid. 
The limit of this approximation is determined by the sound horizon scale, which is given by $c_s/(aH)$~\cite{Sawicki:2015zya, DAmico:2016ntq}.
Given a concrete model, one can check the validity of the quasi-static approximation by using the above expression.

\section{The coefficients of Eqs.~(\ref{Psi F})--(\ref{Q F}) and the $\alpha$ parameters}\label{sec: alphabets}

The coefficients of Eqs.~(\ref{Psi F})--(\ref{Q F}) are given in terms of the $\alpha$ parameters by
\begin{align}
    {\cal F}_{T} 
        & = M^{2}(1+\alpha_{T}),\quad {\cal G}_{T} = M^{2}(1+\alpha_{H}), 
            \notag\\
    A_{0} 
        & = M^{2}\biggl[     \frac{\dot H}{H^{2}} +\frac{3\Omega_{\mathrm{m}}}{2} 
            +\left( 1+\alpha_{M}+\frac{\dot H}{H^{2}}\right)(\alpha_{B}-\alpha_{H})\notag\\
                &\hspace{3em} +\frac{\dot\alpha_{B}-\dot\alpha_{H}}{H}  +(\alpha_{T}-\alpha_{M}) \biggr], 
                    \notag\\
    A_{1} 
        & = M^{2}\left[ \alpha_{H}(1+\alpha_{M})+\alpha_{M} 
            -\alpha_{T} +\frac{\dot\alpha_{H}}{H} \right], 
                \notag\\
    A_{2} 
        & = -M^{2}(\alpha_{B}-\alpha_{H}),
            \notag\\
    A_{3} 
        & = M^{2}\left(\alpha_{M}-\alpha_{T} +\frac{\dot H}{H^{2}}\alpha_{H}\right),
            \notag\\
    B_{0} 
        & = -\frac{M^{2}}{4}\biggl[ \alpha_{G} -3(\alpha_{H}-\alpha_{T})+4\alpha_{B}
            \notag\\
                &\hspace{4em} -\alpha_{M}(2+\alpha_{G} +\alpha_{H}) 
                    -\frac{\dot\alpha_{G}+\dot\alpha_{H}}{H} \biggr],
                        \notag\\
    B_{1} 
        & = \frac{M^{2}}{2}\alpha_{T},\quad 
    B_{2} 
        = \frac{M^{2}}{2}(\alpha_{G}-\alpha_{H}), 
                \notag\\
    C_{0} 
        & = \frac{M^{2}}{4}(\alpha_{G}-\alpha_{H}+\alpha_{T}).
\end{align}

%---------   References   ---------%

\end{document}